%% file: template.tex
\documentclass{Interspeech2024}
\usepackage{color}
\usepackage{tabularx}
\usepackage{multirow}
\usepackage{cite}
\usepackage{graphicx}
\usepackage{subcaption}

\interspeechcameraready

\title{Neural Codec-based Adversarial Sample Detection for Speaker Verification}

\name[affiliation={1^*}]{Xuanjun}{Chen}
\name[affiliation={2^*}]{Jiawei}{Du}
\name[affiliation={1}]{Haibin}{Wu}
\name[affiliation={2}]{Jyh-Shing Roger}{Jang}
\name[affiliation={1}]{Hung-yi}{Lee}

\address{
  $^1$Graduate Institute of Communication Engineering, National Taiwan University\\
  $^2$Department of Computer Science and Information Engineering, National Taiwan University\thanks{\noindent{$^*$} Equal contribution.}}
\email{\{d12942018, r11922185, f07921092, hungyilee\}@ntu.edu.tw, jang@csie.ntu.edu.tw}


\begin{document}

\maketitle

\input{0.abstract}

\input{1.introduction}

\input{2.Related_work}

\input{3.method}

\input{4.experiment}

\input{5.conclusion}

\section{Acknowledgements}
This work was partially supported by the National Science and Technology Council, Taiwan (Grant no. NSTC 112-2634-F-002-005, Advanced Technologies for Designing Trustable AI Services). We also thank the National Center for High-performance Computing (NCHC) of National Applied Research Laboratories (NARLabs) in Taiwan for providing computational and storage resources.


\bibliographystyle{IEEEtran}
\bibliography{mybib}
\end{document}

%% file: 0.abstract.tex
\begin{abstract}

Automatic Speaker Verification (ASV), increasingly used in security-critical applications, faces vulnerabilities from rising adversarial attacks, with few effective defenses available. 
In this paper, we propose a neural codec-based adversarial sample detection method for ASV. 
The approach leverages the codec's ability to discard redundant perturbations and retain essential information.
Specifically, we distinguish between genuine and adversarial samples by comparing ASV score differences between original and re-synthesized audio (by codec models). 
This comprehensive study explores all open-source neural codecs and their variant models for experiments.
The Descript-audio-codec model stands out by delivering the highest detection rate among 15 neural codecs and surpassing seven prior state-of-the-art (SOTA) detection methods.
Note that, our single-model method even outperforms a SOTA ensemble method by a large margin.
\end{abstract}

%% file: 1.introduction.tex
\section{Introduction}

Automatic Speaker Verification (ASV), verifies if a given spoken utterance comes from a specific individual and is widely used in numerous security-sensitive applications, such as banking and access control. 
Deep learning has recently propelled significant progress in ASV, leading to the development of various highly effective ASV models \cite{dehak2010front,kenny12_odyssey,snyder2018x, zhang2022mfa, desplanques2020ecapa, li2020practical}. 
However, ASV remains vulnerable to newly emerged adversarial attacks, presenting significant security risks \cite{kreuk2018fooling, das2020attacker, wu2023defender, jati2021adversarial, villalba2020x, li2020adversarial, marras19_interspeech}.
Adversarial attacks involve creating manipulated inputs, called adversarial samples, to trick machine learning models \cite{szegedy2013intriguing}. 
These samples seem normal to humans but can mislead even sophisticated models. 
This issue affects speech processing areas, such as automatic speech recognition (ASR) \cite{carlini2018audio, qin2019imperceptible}, anti-spoofing for ASV \cite{liu2019adversarial, wu2020defense, wu2020defense_icassp}, and audio-visual active speaker detection \cite{chen2023push}. 
ASV models, including advanced i-vector and x-vector technologies \cite{villalba2020x,li2020adversarial}, are not immune to these attacks. 
Recent studies have been directed towards developing more potent adversarial attacks, focusing on creating universal attacks \cite{marras19_interspeech}, enhancing their transferability in-the-air \cite{li2020practical}, and making them imperceptible \cite{wang2020inaudible}.


Research in adversarial defense for ASV is divided into two main approaches. 
The first approach involves incorporating adversarial samples into training data to create attack-dependent defenses, such as adversarial training \cite{wang2019adversarial} and adversarial sample detection \cite{li2020investigating}, which presupposes knowledge of future attacks — a notion often deemed impractical. 
The second approach focuses on attack-independent defenses, employing strategies like self-supervised learning \cite{wu2021adversarial, wu2021improving}, pre-processing \cite{lan2022adversarial}, voting mechanisms \cite{wu2021voting}, and score-difference detection \cite{wu2022adversarial, chen2023lmd, chen2022masking}, which do not rely on prior knowledge of attacks, offering broader applicability. 
Our proposed codec-based adversarial sample detection belongs to the second method \cite{wu2023scalable}. 


Neural audio codecs \cite{wu2024towards, wu2024codec} compress audio into compact codebook codes, facilitating quicker transmission by quantizing the waveform.
The model consists of an encoder, a lookup codebook, and a decoder. 
The encoder converts audio into frame-wise embeddings, each of which is compared to embeddings in the lookup codebook to find the nearest match, quantizing such frame-wise embeddings to quantized codes. 
The decoder reconstructs the audio by utilizing the embeddings obtained by indexing the quantized codes in the lookup codebook.
Only essential information is retained during quantization, while redundant data is discarded. 
We view adversarial noise as superficial, fragile, and redundant, as it's designed to be naturally imperceptible \cite{szegedy2013intriguing}. 
Through the codec resynthesis process, key audio information is preserved, without hurting the downstream task performance too much for the original audio, while the adversarial noise is discarded.

As a result, this paper introduces a novel approach employing neural codecs to identify adversarial audio in ASV, achieving a performance surpassing previous state-of-the-art methods \cite{wu2022adversarial, wu2023scalable}. 
Please note that the method proposed by \cite{wu2023scalable} is based on an ensemble approach. 
Interestingly, our single-model method outperforms this ensemble method by a large margin.
Specifically, we differentiate between genuine and adversarial samples by comparing ASV score differences between original and re-synthesized audio using codecs. 
This paper investigates all open-source neural codecs, including SpeechTokenizer \cite{zhang2023speechtokenizer}, AcademiCodec \cite{yang2023hifi}, AudioDec \cite{wu2023audiodec}, Descript-audio-codec (DAC) \cite{kumar2023high}, Encodec \cite{defossez2022high}, and FunCodec \cite{du2023funcodec}, for this purpose.
We compare them in a unified setting and reveal that the DAC achieves the best detection rate among 15 different codecs and 3 different attack budgets, outperforming 7 different previous state-of-the-art (SOTA) detection methods. 
DAC codec also causes the least distortion to the genuine samples. 


%% file: 2.Related_work.tex
\section{Background}

\textbf{Automatic speaker verification.}
ASV encompasses the following stages: feature engineering, extracting speaker embeddings, and computing similarity scores.   
Firstly, audio waveforms of speech are transformed into acoustic features, such as spectrograms or filter banks.
Secondly, recent ASV models \cite{kenny12_odyssey, snyder2018x, chung2020defence, zhang2022mfa, desplanques2020ecapa} commonly focus on deriving utterance-level embeddings from the acoustic features.
Lastly, an enrollment utterance is saved in the system, and a testing utterance is input for inference. 
A higher score means a greater likelihood that the testing and enrollment utterances are from the same speaker, and the opposite for a lower score. 
For simplicity, let's refer to the testing utterance as $x_t$ and the enrollment utterance as $x_e$. 
The entire process can be represented as a function $f$:
\begin{equation}
s = f(x_t, x_e)
\end{equation}
where $s$ represents the similarity score between $x_t$ and $x_e$.

\textbf{Adversarial attack.}
Attackers create an adversarial sample by adding nearly imperceptible noise to an original sample. 
This sample is essentially the original one altered slightly by a so-called adversarial perturbation or noise, designed to lead the model to incorrect predictions. 
Assuming the attackers know the ASV system's architecture, parameters, and gradients, along with access to the test utterance $x_t$, they aim to craft an adversarial utterance by identifying the adversarial noise. 
They employ various methods to find this noise, each constituting a different attack algorithm. 
This paper focuses on the Basic Iterative Method (BIM) \cite{kurakin2016adversarial}, which is recognized for its effectiveness and strength. 
The BIM process starts with the attackers setting $x^0_t = x_t$ and then iteratively refining it to produce the adversarial sample as follows:
\begin{multline}
 x^{k+1}_t = \text{clip} \left( x^{k}_t + \alpha \cdot (-1)^I \cdot \text{sign} \left( \nabla_{x^k_t} f(x^k_t, x_e) \right) \right),\\
 for\quad k = 0, 1, ..., K - 1,
\end{multline}
where the update is restricted by the condition ${||x^{k+1}_t - x^k_t||}_\infty \leq \epsilon$, with the clip (.) representing the clipping function to maintain the perturbation within limits set by the attack budget $\epsilon$, where $\epsilon \geq 0 \in \mathbb{R}$. 
The parameter $\alpha$ is the step size, and $I$ takes the value 1 for target trials and 0 for non-target trials. 
The total number of iterations, $K$, is determined by $\lceil \epsilon / \alpha \rceil$, using the ceiling function. 
The final adversarial sample is denoted by $x^k_t$. 
In cases of non-target trials, where the test and enrollment utterances are from different speakers, the BIM aims to fool the ASV system into producing high similarity scores for these utterances, causing it to accept an imposter mistakenly.

\textbf{Neural codec.}
The evolution of neural codec models \cite{wu2024towards,wu2024codec} has seen significant advancements through various innovative implementations to enhance audio quality and efficiency. 
SoundStream \cite{zeghidour2021soundstream} introduces a foundational framework utilizing encoder, quantizer, and decoder modules, leveraging streaming SEANets \cite{tagliasacchi2020seanet} for encoding and decoding, and incorporating a speech enhancement system within its quantizer. 
Encodec \cite{defossez2022high} builds upon this by refining the architecture and integrating advanced techniques such as LSTM layers and Transformer-based models for improved sequence modeling. 
Further developments are marked by AudioDec's \cite{wu2023audiodec} implementation of group convolution for real-time operation and the HiFi-GAN \cite{kong2020hifi} for high-fidelity audio generation. 
AcademiCodec \cite{yang2023hifi} introduces group-residual vector quantization for efficient generation tasks, while SpeechTokenizer \cite{zhang2023speechtokenizer} focuses on a unified speech tokenizer architecture that enhances semantic and acoustic token separation. 
Descript-audio-codec \cite{kumar2023high} stands out for its broad data type fidelity, employing various training techniques to maintain audio quality. 
Finally, FunCodec \cite{du2023funcodec} proposes a frequency-domain approach, demonstrating that incorporating semantic information can significantly improve speech quality, especially at lower bit rates.

%% file: 3.method.tex
\section{The neural codec detection framework}


As shown in Figure \ref{fig:unified_codec}, the proposed neural codec detection framework integrates an ASV and a neural codec system.

\input{Figure_tex/overview}

\subsection{Codec-based detection procedure}

\textbf{Definition of the score difference.}
The overall detection framework is shown in Figure~\ref{fig:unified_codec}, where $x$ represents the input testing utterance and $x'$ denotes the testing utterance after audio codec processing. 
To streamline the description, we exclude the enrollment utterance and the subscript of $x_t$.
We calculate each testing utterance's absolute difference $|s - s'|$. 
For genuine samples, the value of $|s - s'|$ approaches zero. 
In contrast, in the case of adversarial samples, the value of $|s - s'|$ is relatively larger. 
Therefore, we can establish a threshold to differentiate adversarial samples from genuine samples. 
Score variation is denoted as $d=|s - s'|$, and we define $\mathbb{T}_{g}$ as the set of genuine testing utterances, represented as $\mathbb{T}_{g} = \left\{ x_g^{1}, x_g^{2}, \ldots, x_g^{i} \right\}$.
For each genuine testing utterance $x_g^{i}$, the score variation $d_g^{i}$ can be calculated as $d_g^{i} = {|s_g^{i} - {s_g^{i}}'|}$, where $s_g^{i}$ represents the ASV score of $x_g^{i}$ without processing by an audio codec, and ${s_g^{i}}'$ represents the score after processing.

\input{Tables/codec_defake_table}

\textbf{Determine the detection threshold.}
In reality, due to the inability of ASV system designers to determine whether a testing utterance is an adversarial sample, the detection threshold is established solely based on genuine samples. 
Consequently, prior knowledge of adversarial attack algorithms is not required for this detection method.
Additionally, according to the system design requirements from users, we can manually set a false detection rate of genuine data ($ FPR_{given} \in [0,1]$) to meet the sensitivity adjustment requirements of ASV system designers. 
Then we can derive a detection threshold $\tau_{det}$:
\begin{align}
    &FPR_{det}(\tau) = \frac{\vert \{ d_g^{i} > \tau : x_{g}^{i} \in \mathbb{T}_{g} \} \vert}{\vert \mathbb{T}_{g} \vert} \label{eq:det-far} \\
    &\tau_{det} = \{ \tau \in \mathbb{R} : FPR_{det}(\tau) =FPR_{given} \} \label{eq:det-threshold} 
\end{align}
where $|\mathbb{T}_{g}|$ is denoted as the total number of genuine samples in the set $\mathbb{T}_{g}$, and $d_g^{i}$ is derived by $x_g^{i}$ as illustrated in Figure. 1.

\textbf{Inference procedure.}
For this detection method, regardless of whether the input testing utterance is an adversarial sample or a genuine sample, the system will calculate score variation $d$. 
In the cases of $d < \tau_{det}$, the current testing utterance will be labeled as a genuine sample by the detection system, and vice versa. 
In correspondence, we denote the score variation of adversarial samples as $d_a^{i}$. 
According to Eq.~\ref{eq:det-far}, we can derive the detection rate ($DR_{\tau_{det}}$ ) under $\tau_{det}$: 
\begin{align}
    &DR_{\tau_{det}} = \frac{\vert \{ d_{a}^{i} > \tau_{det} : x_{a}^{i} \in \mathbb{T}_{a} \} \vert}{\vert \mathbb{T}_{a} \vert} \label{eq:det-rate} 
\end{align}
where $|\mathbb{T}_{a}|$ is the set of adversarial testing utterances, and $d_a^{i}$ represents the score variation of $x_{a}^{i}$, as illustrated in Figure~\ref{fig:unified_codec}. 
The detection rate ($DR_{\tau_{det}}$) will serve as the primary criterion for evaluating the performance of the audio neural codec to detect adversarial attacks.

\subsection{The rationale behind the codec-based detection}
\label{lab:rationable}

As depicted in Figure \ref{fig:unified_codec}, the architecture of a standard neural codec model is comprised of three key components: an Encoder, a Residual Vector Quantizer (RVQ), and a Decoder. 
The primary objective of the Encoder is to compress speech to quantized codes with key information, and the Decoder will decode the quantized codes back to speech signals.
During quantization, only essential information is preserved, while redundant information is discarded. 
Adversarial noise is considered redundant since it's designed to be naturally imperceptible.
Through the codec resynthesis process, crucial audio information is retained without significantly impairing the downstream task performance for the original audio, while the adversarial noise is eliminated.
Finally, the codec effectively minimizes ASV score degradation for genuine samples and mitigates adversarial noise by focusing on essential audio information and discarding non-essential information (e.g., imperceptible adversarial perturbations). 
The minimal degradation of ASV performance for genuine samples is evidenced by EER presented in the last column of Table~\ref{tab:overall_codec_information}-(b). 
In cases where the input consists of adversarial samples, the codec model endeavors to quantize the waveform into discrete units, effectively neglecting minor perturbations.

%% file: Figure_tex/overview.tex
\begin{figure}[t]
\includegraphics[width=8.0cm]{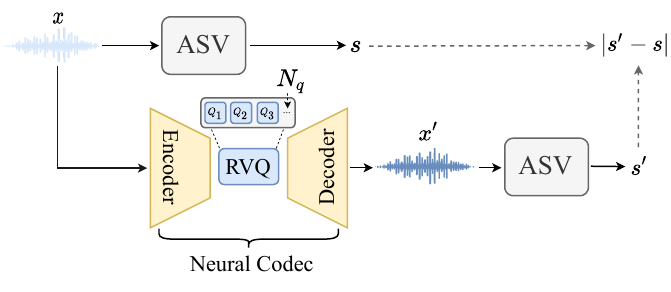}
\vspace{-7mm}
\caption{The neural codec detection framework. $s$ and $s'$ are the ASV scores for $x$ and $x'$. The absolute value $|s - s'|$ between $s$ and $s'$ is for detection.}
\label{fig:unified_codec}
\end{figure}

%% file: Tables/codec_defake_table.tex
\begin{table*}[t]
\centering
\fontsize{8}{10}\selectfont
\caption{Neural codec model comparison.}
\label{tab:overall_codec_information}
\vspace{-2mm}
\begin{tabular}{lccccc|cccc}
  \toprule
  \multicolumn{6}{c}{(a) Codec Information} & \multicolumn{4}{c}{(b) Detection rate} \\
  \midrule
    & \multirow{2}{*}{Codec model} & \multirow{2}{*}{Other Configuration} & \multirow{2}{*}{kbps} & \multirow{2}{*}{$N_q$} & \multirow{2}{*}{SR} & \multicolumn{4}{c}{{\shortstack{Attack budget $\epsilon = 10$}}} \\
  & & & & & & FPR 0.05 & FPR 0.01 & FPR 0.005 & FPR 0.001 \\
  \toprule
  A & SpeechTokenizer \cite{zhang2023speechtokenizer} & 16k & 4 & 8 & 16 & 92.75 & 86.74 & 83.51 & 75.96 \\
  \midrule
  B1 & \multirow{3}{*}{AcademiCodec \cite{yang2023hifi}} & 16k\_320d & 2 & 4 & 16 & 83.17 & 71.37 & 66.09 & 52.43 \\
  B2 & & 16k\_320d\_large\_uni & 2 & 4 & 16 & 84.85 & 73.55 & 67.38 & 52.26 \\
  B3 & & 24k\_320d & 3 & 4 & 24 & 83.32 & 72.45 & 69.34 & 53.40 \\
  \midrule
  C & AudioDec \cite{wu2023audiodec} & 24k\_320d & 6.4 & 8 & 24 & 82.41 & 71.12 & 67.31 & 53.26 \\
  \midrule
  D1 & \multirow{3}{*}{DAC \cite{kumar2023high}} & 16k & 6 & 12 & 16 & \textbf{99.02} & \textbf{97.90} & \textbf{97.42} & 95.73 \\
  D2 & & 24k & 24 & 32 & 24 & 98.10 & 97.23 & 96.85 & \textbf{95.84} \\
  D3 & & 44k & 8 & 9 & 44.1 & 98.19 & 96.39 & 95.50 & 91.82 \\
  \midrule
  E1 & EnCodec \cite{defossez2022high} & 24k & 1.5 & 2 & 24 & 48.52 & 31.83 & 25.63 & 14.87 \\
  \midrule
  F1 & \multirow{6}{*}{FunCodec \cite{du2023funcodec}} & en\_libritts\_16k\_gr1nq32ds320 & 16 & 32 & 16 & 98.08 & 96.32 & 95.23 & 92.39 \\
  F2 & & en\_libritts\_16k\_gr8nq32ds320 & 16 & 32 & 16 & 97.87 & 95.21 & 93.32 & 87.61 \\
  F3 & & en\_libritts\_16k\_nq32ds320 & 16 & 32 & 16 & 95.60 & 90.87 & 87.90 & 78.34 \\
  F4 & & en\_libritts\_16k\_nq32ds640 & 8 & 32 & 16 & 95.80 & 90.94 & 88.22 & 78.69 \\
  F5 & & zh\_en\_16k\_nq32ds320 & 16 & 32 & 16 & 97.42 & 94.89 & 93.16 & 86.36 \\
  F6 & & zh\_en\_16k\_nq32ds640 & 8 & 32 & 16 & 98.03 & 96.08 & 94.80 & 89.88 \\
  \bottomrule
\end{tabular}%
\end{table*}

%% file: 4.experiment.tex
\section{Experiment}
\subsection{Experimental setup}
We adopted a variation of the X-vector system \cite{chung2020defence} as our ASV model. 
The training dataset is VoxCeleb2 \cite{chung2018voxceleb2}, and the evaluation set is VoxCeleb1 \cite{nagrani2017voxceleb}.
Spectrograms are generated using a 25ms Hamming window with a 10ms stride, while 64-dimensional filter banks are utilized as the input features. 
The training process does not incorporate additional data augmentation or voice activity detection techniques. 
Cosine similarity is employed for scoring in the backend. 
The ASV system's performance and the detection performance of adversarial samples are evaluated using the test trials from the VoxCeleb1-O dataset.

\subsection{Experimental results}

\input{Tables/performance_under_diff_attack_budgets}

\textbf{Neural codecs.} As shown in Table ~\ref{tab:overall_codec_information}-(a), we use 15 neural codec variants based on six open-source codec models for codec-based detection methods. 
We first assess the detection rate of different neural codecs for codec-based detection methods in Table~\ref{tab:overall_codec_information}-(b). 
Take the most stringent case with FPR 0.001 in the last column, for example, and we can find that DAC models D1-D3 achieve the best detection performance among 15 codec-based detection methods. 
The detection rate of FunCodec models F1-F6 is around $80\sim90$\% in most cases. Encodec E1 has the worst detection rate with only 14.87\%.
We further compare horizontally and find that DAC and Funcodec are the most effective families.
DAC D1 and FunCodec F1 perform the best in their families, respectively, for most FPR cases. 
We will take D1 and F1 as the most effective models in the following experiments.



\textbf{Attack budgets.}
We compared the codec-based detection methods D1 and F1 with a vocoder-based method \cite{wu2022adversarial} (previous single-model SOTA) using Parallel WaveGAN \cite{yamamoto2020parallel} to assess their performance across different attack budgets in Table~\ref{tab:combined_diff_codec_detection_eer}-(a).
Taking $FPR_{given}=0.001$ as examples, we can observe that D1 and F1 outperform baseline 14.71\% and 7.82\%, respectively. There is a similar trend in different attack budgets and $FPR_{given}$. 
The above experiments show that codec-based detection methods are successful against adversarial noise. 
However, codecs also have side effects that might distort genuine samples.
We further examine whether D1 and F1 can preserve speaker information after codec re-synthesis by evaluating on speaker verification task. 
We try to address this concern by evaluating the re-synthesized audio on the speaker verification task. In Table~\ref{tab:combined_diff_codec_detection_eer}-(b), the results show that DAC preserves better speaker information than F1, the vocoder-based method.


\textbf{SOTA detection methods.}
We compared our codec-based detection method against current SOTA approaches \cite{wu2023scalable} in Table~\ref{tab:combined_diff_codec_detection_eer}-(c).  
To ensure a fair evaluation of our approach, our experimental setup is consistent with that of the SOTA method \cite{wu2023scalable}. 
Our DAC D1 method surpasses the previous SOTA ensemble method (S3) and single methods (S4-S8), demonstrating superior effectiveness with improvements of 0.69\% and 1.16\% under two FPRs, respectively. 
This highlights the efficacy of our approach in detecting adversarial samples.
We conducted a significance test between models S1 and S6, finding the p-value to be $4.62 \times 10^{-114}$. This result indicates a statistically significant improvement of model S1 over S6.


\subsection{Discussion}
\input{Figure_tex/reasonability}

\input{Figure_tex/trade-off_codec}

\textbf{The rationales behind codec-based adversarial detection:}
Figure~\ref{fig:reasonability} presents the score-difference distribution of genuine and adversarial samples across codec models D1, E1, and F1. 
All adversarial samples have a consistently attacked budget with $\epsilon = 10$. 
The $|s' - s|$ refer to testing utterance's score difference. 
The genuine samples for D1 and F1 showcase a steep decline followed by a plateau, which means the score variation of D1 and F1 is extremely small, close to zero.  
In contrast, adversarial samples for D1 and F1 appear to follow a bell-shaped distribution, peaking around 0.3 and 0.4, respectively, and D1's kurtosis is higher than that of F1. 
We can see clear boundaries to distinguish genuine and adversarial distributions for D1 and F1. 
For genuine and adversarial samples of E1, their distributions overlap, which implies that E1 is ineffective in obtaining a threshold to distinguish genuine and adversarial attacks.

\textbf{Limitation.}
The codec-based detection method is helpful against adversarial samples but might cause genuine sample distortion. 
Figure~\ref{fig:codec_trade-off}, with the y-axis showing EER for adversarial data (lower is better for security) and the x-axis for genuine data EER (lower is better for user experience), illustrates this trade-off. 
Ideal performance is in the bottom-left, showing both low adversarial and genuine data EER. 
The performance without codecs is top-left.
This figure shows that no codec perfectly balances reducing noise without affecting genuine data quality.

%% file: Tables/performance_under_diff_attack_budgets.tex
\begin{table}[t]
\centering
\fontsize{8}{10}\selectfont
\caption{The overall codec-based method performance}
\label{tab:combined_diff_codec_detection_eer}
\vspace{-1.5mm}
(a) Detection rate under different attack budget $\epsilon$
\vspace{-1mm}
\begin{tabularx}{0.47\textwidth}{c c c c c}
\toprule
\multirow{2}{*}{$FPR_{given}$} & \multirow{2}{*}{Method} & \multicolumn{3}{c}{Different attack budget $\epsilon$} \\
                               &          & 15 & 10 & 5 \\
\midrule
\multirow{3}{*}{0.05}& DAC D1  &\textbf{99.72} & \textbf{99.02} & \textbf{95.19}  \\
                     & FunCodec F1      & 99.13& 98.08& 92.11  \\
                     & Parallel WaveGAN & 98.82&97.30&89.33 \\
\midrule
\multirow{3}{*}{0.01} & DAC D1   & \textbf{99.32} & \textbf{97.90} & \textbf{90.97} \\
     & FunCodec F1       & 98.30 & 96.32 & 86.31 \\
     & Parallel WaveGAN  & 97.56 & 94.76 & 81.60 \\
\midrule
\multirow{3}{*}{0.005} & DAC D1 & \textbf{99.09} & \textbf{97.42} & \textbf{89.00}  \\
     & FunCodec F1       & 97.77 & 95.23 & 83.61  \\
     & Parallel WaveGAN  & 96.78 & 93.25 & 78.21  \\
\midrule
\multirow{3}{*}{0.001} & DAC D1  & \textbf{98.29} & \textbf{95.23} & \textbf{83.29} \\
     & FunCodec F1       & 96.01 & 92.39 & 76.40 \\
     & Parallel WaveGAN  & 93.89 & 88.60 & 68.58  \\
\bottomrule
\end{tabularx}

\vspace{3mm} 

(b) Adversarial EER under different attack budget $\epsilon$
\setlength{\tabcolsep}{9pt}
\vspace{-1mm}
\begin{tabularx}{0.47\textwidth}{c c c c c}
\toprule
\multirow{2}{*}{Method} & \multicolumn{4}{c}{Different attack budget $\epsilon$} \\
       & 15 & 10 & 5 & 0 \\
\midrule
DAC D1           & \textbf{40.75} & \textbf{29.31} & \textbf{15.84} &  3.32 \\
FunCodec F1      & 66.16 & 50.97 & 27.93 &  3.45 \\
Parallel WaveGAN  & 65.75 & 52.20 & 30.37 &  3.39 \\
None             & 95.65 & 90.57 & 74.05 &  \textbf{2.88} \\
\bottomrule
\end{tabularx}

\vspace{3mm} 
(c) Detection rate of SOTA detection methods
\setlength{\tabcolsep}{8.5pt}
\vspace{-1mm}
\begin{tabularx}{0.47\textwidth}{lc l l}

\toprule
  & \multirow{2}{*}{Method} & \multicolumn{2}{c}{Attack budget $\epsilon=15$} \\    
  &               &  FPR = 0.01   &  FPR = 0.001 \\
\midrule
S1 & DAC D1           & \textbf{99.32} ({\color{red}+0.69}) & \textbf{98.29} ({\color{red}+1.16}) \\
S2 & FunCodec F1      & 98.30 & 96.01  \\
\hline
S3 & Generation+Gaussian & \textbf{98.63} & \textbf{97.13}  \\
S4 & Generation        & 97.62 & 94.71  \\
S5 & Gaussian          & 96.84 & 93.21  \\
S6 & Parallel WaveGAN  & 97.51 & 94.07  \\
S7 & HiFi-GAN          & 94.48 & 79.42  \\
S8 & Encodec           & 95.90 &  90.71 \\
\bottomrule
\end{tabularx}
\end{table}

%% file: Figure_tex/reasonability.tex
\begin{figure}[t]
\centering
\includegraphics[width=7.5cm]{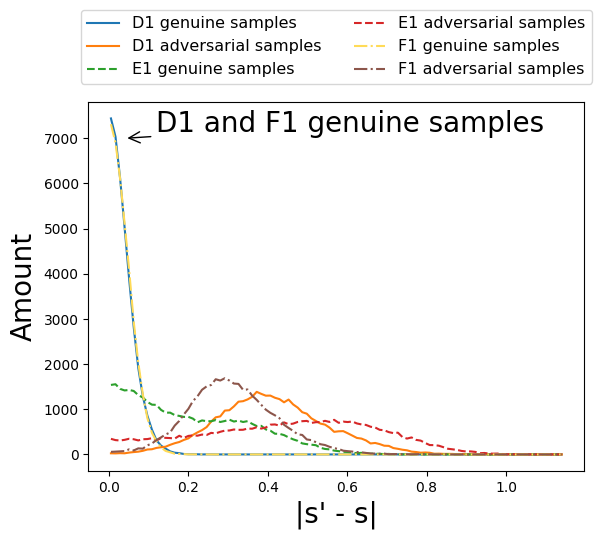}
\vspace{-2mm}
\caption{The score-difference distributions.} 
\label{fig:reasonability}
\end{figure}

%% file: Figure_tex/trade-off_codec.tex
\begin{figure}[t]
\centering
\vspace{-2.5mm}
\includegraphics[width=7.0cm]{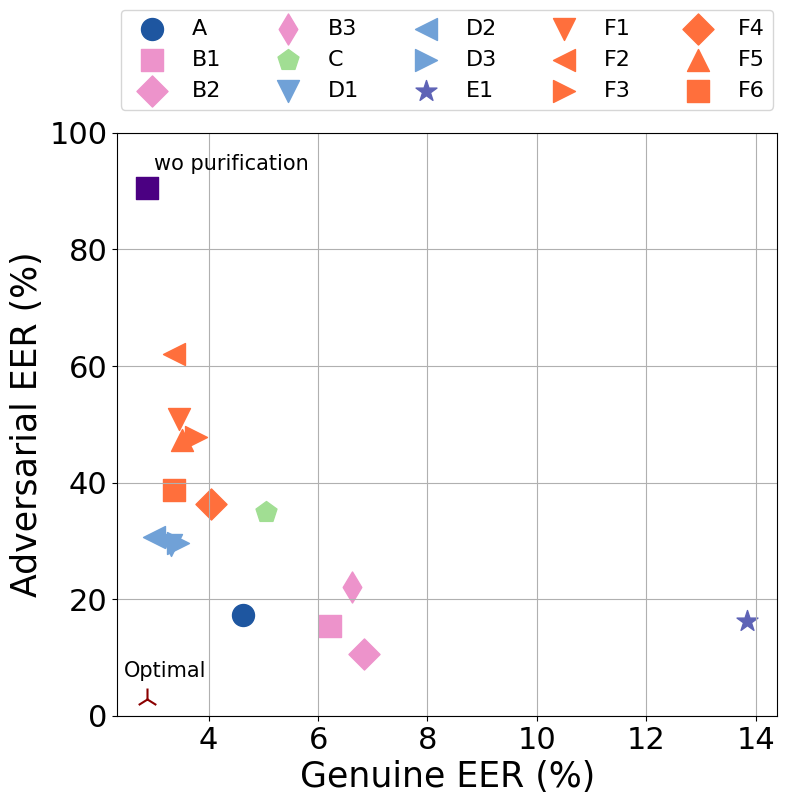}
\vspace{-2mm}
\caption{The trade-off for neural codec purification.}
\label{fig:codec_trade-off}
\end{figure}

%% file: 5.conclusion.tex
\section{Conclusion}

This paper presents a novel neural codec-based detection method for detecting adversarial samples in ASV, with the DAC codec-based detection method showing SOTA performance in detection rate and minimal distortion to genuine samples. 
Our analysis reveals the quantization mechanism of neural codec models is effective in ignoring minor perturbations, particularly for adversarial noise.
Future research will focus on optimizing codec models to achieve an effective balance between genuine and adversarial EERs.